\documentclass[aip,reprint]{revtex4-1}

\usepackage{graphicx}
\usepackage{dcolumn}
\usepackage{bm}
\usepackage{amssymb}
\usepackage[T1]{fontenc}
\usepackage{textcomp}
\begin{document}

\title{Perpendicularly magnetized CoFeB multilayers with tunable interlayer exchange for synthetic ferrimagnets}

\author{P. Pirro}
\altaffiliation{currently at: Fachbereich Physik and Landesforschungszentrum OPTIMAS, Technische Universit\"at
	Kaiserslautern, 67663 Kaiserslautern, Germany}
\affiliation{Institut Jean Lamour, Universit\'e de Lorraine, UMR 7198 CNRS, 54506 Vandoeuvre-l\`es-Nancy, France}
\author{A. Hamadeh}
\affiliation{Institut Jean Lamour, Universit\'e de Lorraine, UMR 7198 CNRS, 54506 Vandoeuvre-l\`es-Nancy, France}
\author{M. Lavanant-Jambert}
\affiliation{Institut Jean Lamour, Universit\'e de Lorraine, UMR 7198 CNRS, 54506 Vandoeuvre-l\`es-Nancy, France}
\author{T. Meyer}
\affiliation{Fachbereich Physik and Landesforschungszentrum OPTIMAS, Technische Universit\"at
Kaiserslautern, 67663 Kaiserslautern, Germany}
\author{B. Tao}
\affiliation{Institut Jean Lamour, Universit\'e de Lorraine, UMR 7198 CNRS, 54506 Vandoeuvre-l\`es-Nancy, France}
\author{E. Rosario}
\affiliation{Institut Jean Lamour, Universit\'e de Lorraine, UMR 7198 CNRS, 54506 Vandoeuvre-l\`es-Nancy, France}
\author{Y. Lu}
\affiliation{Institut Jean Lamour, Universit\'e de Lorraine, UMR 7198 CNRS, 54506 Vandoeuvre-l\`es-Nancy, France}
\author{M. Hehn}
\affiliation{Institut Jean Lamour, Universit\'e de Lorraine, UMR 7198 CNRS, 54506 Vandoeuvre-l\`es-Nancy, France}
\author{S. Mangin}
\affiliation{Institut Jean Lamour, Universit\'e de Lorraine, UMR 7198 CNRS, 54506 Vandoeuvre-l\`es-Nancy, France}
\author{S. Petit Watelot}
\affiliation{Institut Jean Lamour, Universit\'e de Lorraine, UMR 7198 CNRS, 54506 Vandoeuvre-l\`es-Nancy, France}

\date{\today}

\begin{abstract}
A study of the multilayer system MgO/CoFeB(1.1nm)/Ta($t$)/CoFeB(0.8nm)/MgO is presented, where the two CoFeB layers are separated by a Ta interlayer of varying thickness $t$. The magnetization properties deduced from complementary techniques such as superconducting quantum interference magnetometry, ferromagnetic resonance frequency measurements and Brillouin light scattering spectroscopy can be tuned by changing the Ta thickness between $t$=0.25 nm, 0.5 nm and 0.75 nm.
For $t$=0.5 nm, a ferromagnetic coupling is observed, whereas for t=0.75 nm, the antiferromagnetic coupling needed to construct a synthetic ferrimagnet is realized. In the later case, the shape of magnetic domain walls between two ferrimagnetic alignments or between a ferro- and a ferrimagnetic alignment is very different. This behavior can be interpreted as a result of the change in dipolar as well as interlayer exchange energy and domain wall pinning, which is an important conclusion for the realization of data storage devices based on synthetic ferri- and antiferromagnets.

\end{abstract}


\maketitle

\section{Introduction}
 
The demand to develop faster and more reliable magnetic memories with large data storage capabilities has lead to the proposition of the racetrack memory concept \cite{Parkin2008}, where data is stored in form of domain walls in magnetic nanowires. Since these domain walls can be moved via spin torque effects to the magneto-resistive readout elements, this concept allows for a three-dimensional data storage device.
However, dipolar effects like, e.g., the Walker breakdown limit the velocity and storage density of magnetic domain walls in single layer wires.  To overcome this limitation, the use of synthetic ferrimagnets (SFI) with perpendicularly-to-plane magnetized layers has proven to be promising \cite{Parkin2015,Kuteifan2104}. Thus, the development and careful characterization of suitable SFI systems with low magnetic damping to reach high domain wall speeds and high device efficiencies is of paramount interest.

In this context, we present a study of the multilayer system MgO(2.5nm)/Co$_{40}$Fe$_{40}$B$_{20}$(1.1nm )/Ta(t)/Co$_{40}$Fe$_{40}$B$_{20}$(0.8nm)/MgO(2.5nm) (all nominal layer thicknesses), where the two CoFeB layers are separated by a Ta interlayer \cite{Cheng2013,Cheng2012} of varying thickness of $t$=0.25, 0.5 and 0.75 nm. This system is directly grown on a GaAs substrate and can thus easily be integrated into semiconductor based devices, e.g. spin LED structures \cite{Lu2014,Tao2016}. In the investigated layer stack, the hybridization of the 3d orbitals of the transition metals with the O2p orbitals of MgO provides the perpendicular magnetic anisotropy (PMA). To establish PMA, thermal annealing is needed to crystalline the CoFeB starting from the CoFeB/MgO interface. During this process, the diffusing boron is absorbed by the Ta interlayer. In addition, Ta provides an RKKY-like coupling \cite{Parkin1991,Cheng2012} which is needed to create a SFI.

Using a superconducting quantum interference device SQUID, magneto-optical Kerr effect microscopy (MOKE), inductive ferromagnetic resonance frequency (FMR) measurements and Brillouin light scattering spectroscopy (BLS), the magnetic properties of the systems have been investigated. The experimental results are consistently reproduced by a simple macrospin model. The interlayer exchange coupling as well as the magnetic properties of the two individual CoFeB layers vary strongly with the Ta thickness $t$. For $t$=0.75 nm, a SFI with an antiferromagnetic (AFM) exchange coupling is observed and the nucleation of domain walls in between the two ferrimagnetic configurations as well as in between the ferri- and the ferromagnetic (FM) configuration are realized.

\section{Sample preparation}

The investigated multilayer stacks are deposited on GaAs substrates which are first desorbed in a MBE chamber at 300$^\circ$C by monitoring reflection high energy electron diffraction patterns to remove the As capping layer, which is used to passivate the surface of the substrate. Then, the samples are transferred to the sputtering chamber without breaking the vacuum to deposit the multilayer at room temperature. The magnetization of the samples prior to an annealing is in-plane, as usual for Ta/CoFeB/MgO systems. To establish PMA, the samples were annealed at 250$^\circ$C for 3 mins in a rapid thermal annealing system (for details, see \cite{Tao2016}. Prior studies of a similar system have show that a magnetically dead layer can be formed at the Ta/CoFeB interface \cite{Jang2011}. According to \cite{Tao2016}, the effective magnetic CoFeB layer thickness $d$ is significantly reduced compared to the nominally deposited thickness if $t \ge 0.5$ nm. The estimated values for $d$ based on the findings in \cite{Tao2016} are given in Table~\ref{parameters}. 

\section{SQUID measurements}

\begin{figure}[h]
\begin{center}
\scalebox{1}{\includegraphics[width=0.5 \textwidth ]{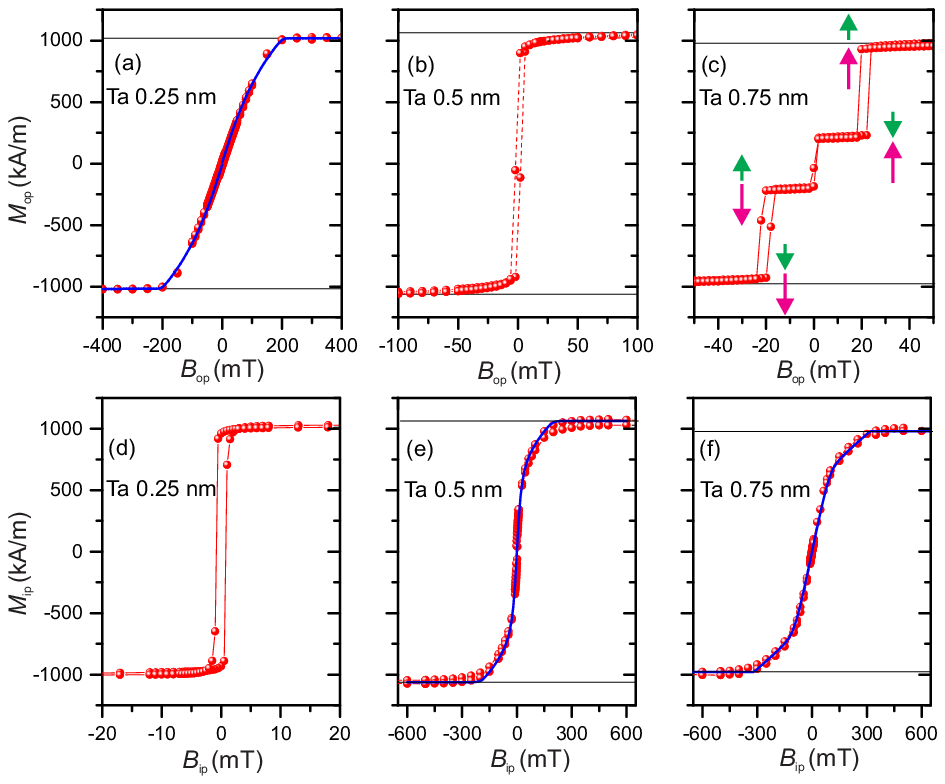}}
\end{center}
\caption{\label{SQUID}(color online) Magnetization as a function of the applied field along the out-of-plane plane direction  (a-c) and along the in-plane direction (d-f) for CoFeB(1.1\,nm)/Ta($t$)/CoFeB(0.8\,nm) with $t=0.25$\,nm (a,d), $t=0.5$\,nm (b,e) and $t=0.75$\,nm (c,f). (c) shows that for $t$=0.75\,nm, the magnetization of both layers is out of plane with a three step hysteresis loop indicating an AFM coupling with a direct transition between the two ferrimagnetic states.}
\end{figure}

Figure \ref{SQUID} shows the magnetization along the field axis as a function of an applied out-of-plane (Fig.\ref{SQUID} (a)-(c)) and in-plane field (Fig.\ref{SQUID} (d)-(e)) as obtained from SQUID measurements at 300 K. The sample with $t$=0.25 nm of Ta shows an in-plane easy axis with a small coercive field of approximately 1 mT and a hysteresis similar to a single layer, thus the two layers of CoFeB seem to be strongly FM coupled, either by strong RKKY-like exchange or by a direct coupling due to a discontinuous Ta layer. The PMA of this system is not sufficient to overcome the demagnetization field and to consequently orient the magnetization out of plane. For $t$=0.5 nm, a single hysteresis loop with small coercivity for both field directions is observed, indicating that either one of the layers has an in-plane easy axis  and the other one an out-of-plane easy axis, or a multidomain state is occurring in remanence (2 mT coercivity field for both directions, not visible in Fig. 1 for the in-plane field due to the scaling). Since the hysteresis loops are both centered around zero field, we conclude that the coupling between the layers is FM. For $t$=0.75 nm, a clear three-step hysteresis is observed for the out-of-plane field (see also Fig.~\ref{MOKE}(a)), whereas for a field in plane a continuous transition without any detectable hysteresis is revealed. Thus, in this case, both layers have an out-of-plane easy axis and are AFM coupled, which leads to the two additional hysteresis loops centered around $B_\mathrm{RKKY}=\pm$20\,mT.

To achieve a more quantitative characterization, the values of the saturation magnetization ${M}_s$ from the SQUID measurements have been extracted. To calculate ${M}_s$ from the measured magnetic moment per area, the effective magnetic thicknesses $d$ given in Table \ref{parameters} have been used.
To  For $t$=0.25 nm,  a saturation magnetization averaged over both layers of ${M}_s=$1019 kA/m is measured, whereas for  $t$=0.5\,nm, ${M}_s$= 1063 kA/m is obtained. In the case of the AFM coupling for $t$=0.75\,nm, assuming that the smaller magnetic moment can be attributed to the thinner layer, a saturation magnetization of  $M_s^1=$1006\,kA/m for the lower layer and $M_s^2=$973\,kA/m for the upper layer (thus averaged ${M}_s=$993 kA/m) can be deduced. 

\section{Macrospin modeling}
To access further material parameters, we model the total energy $E_\mathrm{tot}$ of the system including Zeeman energy $E_\mathrm{Zee}$, demagnetization energy $E_\mathrm{DM}$, uniaxial magnetocrystalline anisotropy energy $E_\mathrm{Ani}$ with the anisotropy easy axis  $\mathbf{u}$ along the out-of-plane direction and the interlayer exchange energy $E_\mathrm{RKKY}$ between the two layers: 
\begin{eqnarray}
E^\mathrm{tot}&=&\sum_{\mathrm{layer} }^{i=1,2}[E^\mathrm{Zee}(\mathbf{M}_i)+E^\mathrm{DM}(\mathbf{M}_i)+E^\mathrm{Ani}(\mathbf{M}_i)] \nonumber\\ 
&+&E^\mathrm{RKKY}(\mathbf{M}_1,\mathbf{M}_2)
\end{eqnarray}

\begin{eqnarray}
E^\mathrm{Zee}(\mathbf{M_i})&=&-\mathbf{M_i} \cdot \mathbf{B} \cdot V_\mathrm{i} \\
E^\mathrm{DM}(\mathbf{M_i})& =&-\frac{\mu_0}{2} (\mathbf{M_i} \cdot\mathbf{u})^2 \cdot  V_\mathrm{i}\\
E^\mathrm{Ani}(\mathbf{M_i})&=&- (\mathbf{m} \cdot \mathbf{u})^2\cdot  K_\mathrm{i}^s \cdot  S \\
E^\mathrm{RKKY}(\mathbf{M_i})&=&- (\mathbf{m_1} \cdot \mathbf{m_2})\cdot  J \cdot  S 
\end{eqnarray}
with the normalized magnetization $\mathbf{m}=\frac{\mathbf{M}}{M_s}$, the out-of-plane unit vector $\mathbf{u}$, the interlayer exchange coupling $J$ and the volume $V_i=S \cdot d_\mathrm{i}$, which depends on the surface $S$ and effective thickness $d_\mathrm{i}$ of the individual CoFeB layer. In the case of AFM coupling ($t$=0.75\,nm), the switching from the ferromagnetic to the ferrimagnetic states takes place at the field $B_\mathrm{RKKY}$, so we can conclude that at this field $E_\mathrm{Zee}=E_\mathrm{RKKY}$, which leads to the estimate of the exchange constant $J=-M_2 \cdot B_\mathrm{RKKY} \cdot d_2\approx- 0.011$  mJ/m$^2$, which is on the same order of magnitude than $J$ of Ref.~\cite{Parkin1991,Cheng2012}. Furthermore, we reconstructed the hard axis SQUID loops by numerically minimizing $E_\mathrm{tot}(\mathbf{M}_1,\mathbf{M}_2)$ as a function of the applied field and subsequently extracting the average magnetization along the field (blue curves in Fig.~\ref{SQUID}(a),(e) and (f)). From this reconstructions, the magnetocrystalline surface anisotropy constants $K_\mathrm{i}^s$ have been obtained which are listed in Table~\ref{parameters}.

\begin{table}
\centering
\resizebox{0.35 \textwidth}{!}{%
\begin{tabular}{c||c|c|c|c|c|c|c|}
$t$ &$d_1$ & $d_2$ &$M_1$ & $M_2$ & $K_\mathrm{1}^s$ & $K_\mathrm{2}^s$  & $J$ \\
\hline
(nm) & \multicolumn{2}{c|}{(nm)} & \multicolumn{2}{c|}{(kA/m)}&\multicolumn{3}{c|}{(mJ/m$^2$)} \\
\hline
0.25  & 1.1 & 0.8 & 1019 & 1019 & 0.54 & 0.54 & $\ge$ 0.4\\
\hline
0.5  & 0.975 & 0.675 & 1063 & 1063  & 0.69 & 0.565 & 0.045\\
\hline
0.75 & 0.84 & 0.54 & 1006 & 973 & 0.58 & 0.385 & -0.011\\
\end{tabular}}
\caption{\label{parameters} Overview of the obtained material parameters. The values for $M$ and $K^s$ as well as the value for the interfacial exchange coupling $J$ for $t$=0.75 nm have been obtained from the SQUID measurements, whereas $J$ in the case of $t$=0.25nm and 0.5 nm are estimates based on the FMR frequency measurements. The effective magnetic thickness $d$ depends on the Ta interlayer thickness $t$. If a distinction between the two layers has not been possible, an average value is given.}

\end{table}

\section{FMR experiments and modeling}

\begin{figure}[h]
\begin{center}
\scalebox{1}{\includegraphics[width=0.5 \textwidth ]{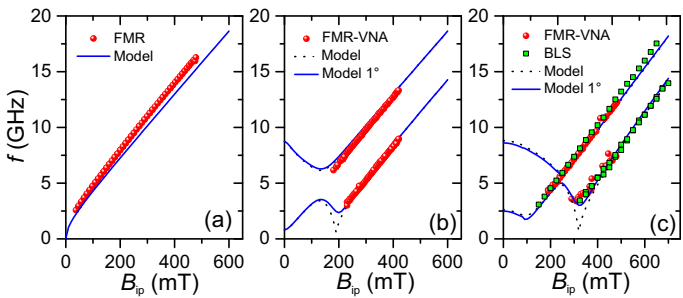}  }
\end{center}
\caption{\label{FMR}(color online) Measurements of the ferromagnetic resonance frequencies (red dots) together with the eigenfrequencies predicted by the macrospin model (blue and black lines) using the material parameters presented in Table~\ref{parameters} obtained from SQUID. Black curves represent the macrospin model for a perfect in-plane field whereas blue curves assume a small out-of-plane tilt of 1$^\circ$.}
\end{figure}

To check the validity of the obtained material constants and to estimate the exchange coupling for the thicknesses $t$ with FM coupling, measurements of the frequencies of the small angle precession eigenmodes have been performed (see Fig.~\ref{FMR}). The external magnetic field has been applied along the in-plane axis and the eigenmode frequencies have been obtained by standard FMR measurements using a vector network analyzer (VNA). In addition, for the case of AFM coupling $t$=0.75 nm, the thermal spin-wave spectrum with a wave vector $k \rightarrow 0$ equivalent to the ferromagnetic resonances has been acquired by Brillouin light scattering spectroscopy \cite{Sebastian2015,Harzer1991}. 

To compare the experimental results with our macrospin model, we transform all energy expressions in a new base given by the equilibrium magnetizations $\mathbf{m}_1^e$ and $\mathbf{m}_2^e$. In this base, we calculate the effective fields 
\begin{equation}
\mathbf{B}^\mathrm{eff}_i=\frac{-1}{V_i \cdot M_i} \nabla_{\mathbf{m}_i^e} E^\mathrm{tot}_i (\mathbf{m}_1^e,\mathbf{m}_2^e)
\end{equation}
and subsequently linearize the two coupled Landau-Lifshitz equations:
\begin{equation}
\frac{d \mathbf{m}^e_1}{dt}=-\gamma \mathbf{m}_1^e \times \mathbf{B}_1^\mathrm{eff}\ \ , \ \frac{d \mathbf{m}^e_2}{dt}=-\gamma \mathbf{m}_2^e \times \mathbf{B}_2^\mathrm{eff}
\end{equation}
with $m^e_z \rightarrow$ 1 (external field axis along $z$) and solve the system numerically as an eigenvalue problem. The frequencies of two modes are obtained, which correspond to the in-phase and out-of-phase oscillations of the two layers when assuming two identical layers. 

For $t$=0.25 nm (Fig.~\ref{FMR}(a)), only one mode within the experimental accessible range from 1 to 20 GHz has been found, again indicating a strong coupling  between the layers. The blue curve shows the frequency evolution predicted by the macrospin model when using the parameters obtained from the SQUID measurement with an exchange coupling $J = 0.4$ mJ/m$^2$. This is the minimal value of $J$ needed to shift the second mode above 20 GHz and thus constitutes the lower boundary for the coupling. A good agreement between the model and the experiment is observed and the small deviation is probably due to a residual misalignment of the applied fields. 

For $t$=0.5 nm (Fig.~\ref{FMR}(b)), two modes with a frequency spacing of about 5 GHz have been observed. Using an exchange coupling of $J$=0.045 mJ/m$^2$ and the parameters obtained from the SQUID, the macrospin model has been evaluated for a perfect in-plane alignment of the external field (dotted black curve) and for a small out-of-plane tilt of the external field of 1$^\circ$ (continuous blue line). The tilt of the field influences especially the local frequency minimum close to 200 mT. Since the trend for the tilted field reproduces the experimentally obtained values more accurately, a slight misalignment from the in-plane axis during the FMR experiment can be concluded. No clear susceptibility peaks have been observed for fields below the in-plane saturation field of 200 mT indicating an inhomogeneous magnetization state which cannot be described by the macrospin model. 

For $t$=0.75 nm (Fig.~\ref{FMR}(c)), again two modes have been observed. Here, with the macrospin model, we only use values obtained from the SQUID measurements. The model nicely reproduces the experimental trend if we allow for a 1$^\circ$ misalignment of the external applied field (continuous blue curve). A perfect alignment of the field along the in-plane direction would again lead to a pronounced frequency minimum, which has not been observed (dotted black curve). Due to the smaller interlayer coupling $J$, the frequency gap which is occurring at about 250 mT with an opening of about 800 MHz is much smaller than in the $t$=0.5 nm case. Due to the comparably large frequency linewidth of more than 1 GHz, the gap could not be observed in the FMR and BLS experiments. Since the frequency linewidth shows no systematic frequency dependence, the underlying broadening mechanism is probably arising from an inhomogeneous distribution of the material parameters.  

The good general agreement between the experimentally obtained frequencies and the calculated eigenfrequencies based on the material parameters of the SQUID measurements demonstrates that a comprehensive and self-consistent characterization of the system has been achieved. 

\section{MOKE microscopy}
\begin{figure}[]
\begin{center}
\scalebox{1}{\includegraphics[width=0.5 \textwidth ]{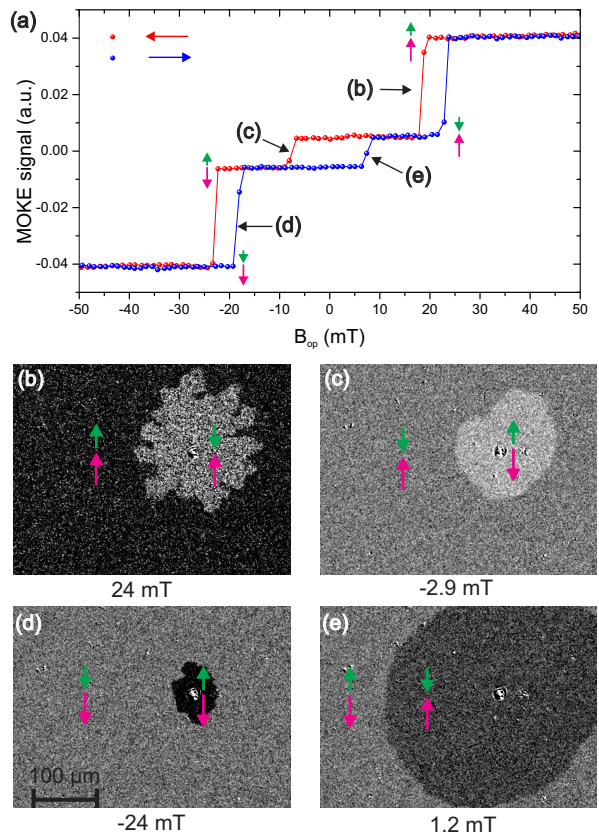} }
\end{center}
\caption{\label{MOKE}(color online) (a) Hysteresis loop obtained from MOKE microscopy for AFM coupling in CoFeB(1.1nm)/Ta(0.75nm)/CoFeB(0.8nm). The arrows indicate the alignment of the two layers for the four different levels of the MOKE signal. (b)-(e) MOKE microscopy images of different domain configurations which have been initially nucleated at the transition fields marked in (a) and imaged at the indicated field values.}
\end{figure}

In the following, we will address the possible domain configurations in the case of the AFM coupling of the  layers ($t$=0.75 nm), since this configuration is especially promising for future high density storage elements \cite{Parkin2015}. The nucleation of DWs was studied experimentally in the presence of an out-of-plane magnetic field. Figure \ref{MOKE} presents magneto-optical Kerr microscopy (MOKE) hysteresis loops and images showing the nucleation of a domain in the different configurations of the two magnetizations in the thinner upper and thicker lower magnetic layer. In Fig.~\ref{MOKE}(a), the magnetic configurations are depicted for the different levels of the MOKE signal by a pair of arrows. Three jumps are observed in the hysteresis loop, similar to the observations using SQUID. Starting from positive fields, the first step (marked by (b) in Fig.~\ref{MOKE}(a)) corresponds to the switching of the upper thinner layer to minimize the interlayer exchange energy, which becomes more important than the Zeeman energy at this field. Thus, a transition from the parallel (P) to the antiparallel (AP) state occurs. The second step (marked by (c)) corresponds to the switching of both layer's magnetization, so a transition between the two AP states is taking place. Here, the Zeeman energy is reduced by aligning the residual magnetic moment of the AP state along the external field, while the interlayer exchange energy is unaffected. With further decreasing field, the Zeeman energy reaches again the value of the interlayer exchange and a switching to the P state occurs (around -25 mT in Fig.~\ref{MOKE}(a)).
In contrast to the SQUID measurement, a minor hysteresis loop around zero field with a clear separation between the two AP states is obtained using MOKE. We attribute this behavior to the different time scales of the measurements: MOKE is significantly faster, so the observed coercivity increases as the system has less time to undergo a thermal activated switching.

Figures \ref{MOKE}(b)-(e) present MOKE images after the nucleation of domains at different external fields around a defect in the thin film. The nucleation takes place at the transitions indicated in Fig.~\ref{MOKE}(a), whereas the fields where the domains are stabilized and imaged are indicated below the pictures.
The MOKE contrast is determined by the net magnetization direction of the domains averaged over the two layers. The orientation of the magnetization is again indicated by arrows.

Two magnetic domains are depicted in Figure \ref{MOKE}(c), each of which includes MOKE contributions from the upper and lower magnetic layer. In the bright domain, the magnetization is oriented down in the lower layer and up in the upper layer. In the darker surroundings, the magnetic orientations of the two layers are both inverted. Consequently, a domain wall is formed between the two AP states which can be considered as a domain wall in a synthetic ferrimagnet. In Fig.~\ref{MOKE}(e), the role of the domain around the defect and the surroundings is just inverted due to the different sign of the nucleation field. For higher fields, the domain configurations shown in Fig.\ref{MOKE}(b) and (d) are stabilized in the center of the outer hysteresis loops and show both a domain with AP alignment of the two layers surrounded by an area with P alignment. Thus, depending on the external field, two different kinds of domain walls can be nucleated in this system. 

Finally, it is clear that the shape of the domain depends on the configuration between the upper and lower magnetic layer.  When the transition occurs between the P and AP orientation (Fig.~\ref{MOKE}(b) and (d)), the domain wall shows a zigzag shape. For the transitions between two AP orientations of the layers (Fig.~\ref{MOKE}(c) and (e)), the domain grows as a circle. The difference of these shapes might be attributed to a combination of different effects which all can lead to distinct domain wall dynamics for the two configurations: first, in the P to AP transition, a domain wall is present only in the upper layer. Thus, it could experience different pinning compared to the domain wall between the two AP states, which is also present in the lower layer. In addition, the dipolar stray fields of a domain with P configuration are much stronger than those generated in the AP configuration, which leads to different contributions of dipolar and interlayer exchange energy in the two cases.

To conclude, we have demonstrated that CoFeB/Ta/CoFeB multilayers grown on GaAs/MgO present strong perpendicular magnetocrystalline anisotropy. The magnetization properties could be tuned by changing the Ta thickness. A 0.5 nm Ta thickness leads to a ferromagnetic coupling whereas a 0.75 nm Ta thickness leads to a synthetic ferrimagnetic bilayer with an antiferromagnetic interlayer coupling. In the later case, the shape of magnetic domain walls between two ferrimagnetic alignments and between a ferromagnetic and a ferrimagnetic alignment is very different, possibly due to the change in dipolar and interlayer exchange energy as well as domain wall pinning. Our results show that with the proper interlayer thickness, the CoFeB/Ta/CoFeB  system is a promising candidate for the realization of data storage devices based on synthetic ferrimagnets. 

\begin{acknowledgments}
This work was supported by the ANR-NSF Project, ANR-13-IS04-0008-01, COMAG by the ANR-Labcom Project LSTNM and by the Universit\'e de la Grande Region (UniGR  funded P.~Pirro Post-Doc). Y. Lu also acknowledges the support by the joint French National Research Agency (ANR)-National Natural Science Foundation of China (NSFC) SISTER project (Grants No. ANR-11-IS10-0001 and No. NNSFC 61161130527) and ENSEMBLE project (Grants No. ANR-14-0028-01 and No. NNSFC 61411136001). Experiments were performed using equipment from the TUBE - Daum funded by FEDER (EU), ANR, the Region Lorraine, and Grand Nancy.
\end{acknowledgments}

\end{document}